\documentclass[aps,reprint,amsmath,amssymb,bibnotes,superscriptaddress]{revtex4-1}

\usepackage{amsfonts,bbold,amsmath,amssymb,graphicx,epstopdf,verbatim}
\usepackage{xcolor}

\begin{document}
 
\title{Spatiotemporal scaling of two-dimensional nonequilibrium exciton-polariton systems with weak interactions}

\author{Quanyu Mei}
\affiliation{Department of Physics, Shanghai Normal University, Guilin Road 100, 200234 Shanghai, China}
\author{Kai Ji}
\email{kji@shnu.edu.cn}
\affiliation{Department of Physics, Shanghai Normal University, Guilin Road 100, 200234 Shanghai, China}
\author{Michiel Wouters}
\affiliation{TQC, Universiteit Antwerpen, Universiteitsplein 1, B-2610 Antwerpen, Belgium}

\date{\today}

\begin{abstract}
We perform a numerical study on the two-dimensional nonequilibrium exciton-polariton systems driven by incoherent pumping based on the stochastic generalized Gross-Pitaevskii equation.
We calculate the density fluctuation, coherence function, and scaling function.
It is found that the correlations at short range agree with the Bogoliubov linear theory.
While at large distance, both static and dynamic correlations are characterized by the nonlinear scaling behaviors of Kardar-Parisi-Zhang (KPZ) universality class, especially when the interaction is weak.
In this regime, scaling analyses are crucial to capture the universal KPZ scaling features.
In addition, the interaction between vortices is modified in the strong KPZ regime and leads to complex nonequilibrium vortex patterns.
\end{abstract}
\maketitle

\section{Introduction}

Nonequilibrium phase transitions have been extensively studied in different fields of natural sciences, as well as in economics and social sciences \cite{Odor-2004, Henkel-2010, Chou-2011}.
For these studies, critical phenomena, scaling and universality are at the center of much interest.
The concept of universality is introduced to classify different systems according to their common critical indices close to a transition point \cite{Stanley-1999}.
In spite of the well-established theory of equilibrium phase transitions, a widely accepted framework for nonequilibrium systems remains elusive.
Nonequilibrium systems may exhibit a variety of new universality classes, such as the genuine, dynamical universality classes, which have no counterparts in equilibrium systems (for a detailed classification, see Ref.~\cite{Odor-2004}).
Recently, renormalization group theory and numerical simulations on the exciton-polariton (EP) condensates in semiconductor microcavities demonstrate that the long-wavelength effective description of the EP systems can be mapped to a compact Kardar-Parisi-Zhang (KPZ) equation \cite{Kardar-1986}, and its long-wavelength dynamics is in the KPZ class, a paradigmatic model for a nonequilibrium phase transition \cite{Halpin-Healy-1995, Takeuchi-2018}.
Hence, the EP systems turns out to be a new platform to investigate the universal nonequilibrium scaling laws \cite{Gladilin-2014, Ji-2015, Altman-2015, He-2017}.

The condensation of EPs is achieved in semiconductor microcavities under nonequilibrium conditions \cite{Kasprzak-2006, Deng-2010}.
This feature is owed to the dynamic balance between the losses and pumping of the cavity subjected to a short lifetime of EP, normally ranging from a few to 100 ps \cite{Carusotto-2013}.
As a typical two-dimensional (2D) system of Bose gas, the EP system develops a quasi-long-range order with a power-law decay in its spacial coherence as the pumping strength increases \cite{Deng-2007, Roumpos-2012, Nitsche-2014}.

However, in the case of incoherent pumping \cite{Kasprzak-2006, Deng-2010}, this quasi-ordered state is found only limited to an intermediate-length-scale due to the KPZ nonlinearity \cite{Altman-2015}.
The standard dynamic equation of EP system, known as the generalized Gross-Pitaevskii equation (GGPE), is domianted by the KPZ physics in the long-wavelength regime, where the nonequilibrium phase fluctuations generated by the drive translate into the nonlinear terms of the KPZ phase equation \cite{Gladilin-2014, Ji-2015, Altman-2015, He-2017, Squizzato-2018}.
As a result, an effective equilibrium cannot be established.
The nonequilibrium nature of the phase fluctuations inevitably destroys the condensate at long scales, which ends up with a stretched exponential decay of the condensate correlations.

Further investigations discover that because of the KPZ nonlinearity, the attraction between unlike vortices is exponentially screened or even become repulsive at large distance showing complex vortex patterns, which is in contrast to the entropy-driven unbinding mechanism for the equilibirum vortices \cite{Wachtel-2016, Gladilin-2017, Gladilin-2019a, Gladilin-2019b}.
Hence, away from equilibrium, the vortices unbind at any finite temperature.
On the contrary, when the nonequilibirum drive is turned off or a strong anisotropy is imposed, the KPZ nonlinearity is reduced and the conventional Berezinskii-Kosterlitz-Thouless (BKT) phase transition \cite{Berezinskii-1972, Kosterlitz-1973, Minnhagen-1987} recovers \cite{Sieberer-2018}.
Such an impact on the form of the vortex interaction is similar to that of driven vortex lattice in the context of the complex Ginzburg-Landau equation \cite{Faller-1998, Aranson-1998a, Aranson-1998b, Faller-1999}.
Very recently, the equilibrium limit of EP system has been reached in experiments owing to high-quality samples with long quasi-particle lifetimes \cite{Sun-2017, Caputo-2018}.
On eliminating the nonequilibrium modulation, vortex states are stabilized, and quasi-long-range order emerges with algebraic decay of coherence observed in both spatial and temporal domains \cite{Caputo-2018}.

An alternative pumping scheme is the coherent pumping, that is to drive the microcavity polaritons into the optical parametric oscillator (OPO) regime \cite{Stevenson-2000, Baumberg-2000}.
Here,  the KPZ effect is also relevant and the fluctuation is manifested as the relative phase of the signal and idler modes, which acts as the Goldstone mode that determines the physics at large scales.
In OPO systems, the KPZ nonlinearity and the degree of anisotropy can be tuned over a wide range of values by changing the driving strength, the pump wave vector of the laser, and the detuning between photons and excitons.
For weak anisotropy, the OPO system falls into the KPZ universality class similar to the incoherently pumped polaritons mentioned above.
While for strong anisotropy, which is inaccessible to the current incoherent pumping experiment, the system is described at large scales by a dynamical XY model.
Thus an algebraic superfluidity order as well as the BKT transition become realizable in the strongly anisotropic OPO system \cite{Dagvadorj-2015, Zamora-2017}.
Within such a scenario, Comaron et al. theoretically study the phase ordering after an infinitely rapid quench across the critical region \cite{Comaron-2018}.
It is confirmed that the system features diffusive dynamics of topological defects and satisfies the dynamical scaling hypothesis.

Although the dominant role of KPZ nonliearity at the large distance has been elaborated in the literature, its impact on the spatiotemporal coherence of EP systems is less understood, especially for the incoherently pumped microcavity polaritons.
In the present work, we close this gap by performing a numerical study on the spatiotemporal coherence of 2D nonequilibrium EP systems in the incoherent pumping regime.
The scaling function is extracted from the first order coherence function, which displays characteristic KPZ scaling behaviors particularly for weak interaction.
This paper is organized as follows.
In Sec. II, we introduce the model and Bogoliubov linear formalism to get a rough insight into the spatiotemporal coherence.
Then we describe our schemes of numerical simulation which adequately takes into account the nonlinear fluctuations.
In Sec. III, the emergence of KPZ nonlinearity is discussed in details together with numerical results on density fluctuation, coherence function and scaling function.
Our conclusions are finally drawn in Sec. IV.

\section{Theory and methods \label{sec:the}}

We starts with the GGPE plus a noise term \cite{Chiocchetta-2013} (we set $\hbar = k_B = 1$),
\begin{eqnarray}    \label{ggpe}
i {d \psi({\bf r},t) \over dt} &=& \left[ - {\nabla^2 \over 2m} + g |\psi|^2 +i \left( {P_0 \over 1+|\psi|^2/n_s} - \gamma \right) \right] \psi({\bf r},t)
  \nonumber \\
  & & + {dW({\bf r},t) \over dt} ,
\end{eqnarray}
where $\psi ({\bf r},t)$ is the classical field wave function, $g$ the interaction strength, $P_0$ the pump strength, $n_s$ the saturation density, and $\gamma$ the damping rate.
The last term of Eq.~\eqref{ggpe} represents the effect of random noise.
Its correlations are taken to be Gaussian and uncorrelated in both space and time:
\begin{eqnarray}
\langle dW({\bf r},t) dW^*({\bf r}',t') \rangle = 2D \delta({\bf r}-{\bf r}') \delta(t-t') dt dt' ,
\end{eqnarray}
where $D$ is the diffusion coefficient, proportional to the strength of noise.

In the absence of noise, the steady state density of GGPE is $n_0 \equiv |\psi_0|^2 = n_s(P_0/\gamma-1)$ when the pumping exceeds the losses ($P_0 > \gamma$).
If the Bogoliubov theory of quantum fluids is valid, we can linearize the macroscopic wave function near the steady state as
$\psi({\bf r},t) = \left[ \psi_0 + \delta \psi({\bf r},t) \right] e^{-i \mu t}$,
with $\delta \psi({\bf r},t)$ the fluctuation of field and $\mu = g n_0$ the oscillation frequency determined by the self-interaction energy of the bosons.
In the presence of noise, the linearization of the GGPE leads to an equation of motion for the fluctuation in the Fourier space \cite{Chiocchetta-2013}
\begin{eqnarray}    \label{g1_kt2}
i \left( \begin{array}{c} d \delta \psi_{\bf k} \\ d \delta \psi^*_{-{\bf k}} \end{array} \right) =
\mathcal{L} \left( \begin{array}{c} \delta \psi_{\bf k} \\ \delta \psi^*_{-{\bf k}} \end{array} \right) dt
+ \left( \begin{array}{c} d W_{\bf k} \\ -d W^*_{-{\bf k}} \end{array} \right) ,
\end{eqnarray}
with $\mathcal{L}$ the Bogoliubov matrix defined as
\begin{eqnarray}
\mathcal{L} = \left( \begin{array}{cc} \epsilon_{\bf k} + \mu - i \Gamma & \mu - i \Gamma \\ - \mu - i \Gamma & - \epsilon_{\bf k} - \mu -i \Gamma \end{array} \right) ,
\end{eqnarray}
where $\epsilon_{\bf k}=k^2/(2m)$ is the kinetic energy of the bosons and $\Gamma=\gamma(P_0-\gamma)/P_0$ the dressed damping rate of the fluctuation field.
After diagonalizing $\mathcal{L}$, one gets two branches of eigen energies, $\lambda^{\pm}_{\bf k} = -i \Gamma \pm i \omega_{\bf k}$, where $\omega_{\bf k} = \sqrt{|\Gamma^2 - E_{\bf k}^2|}$, and $E_{\bf k}=\sqrt{\epsilon_{\bf k} (\epsilon_{\bf k}+2\mu)}$ is the energy dispersion of standard Bogoliubov mode.

The physical quantity we are interested in is the first order coherence function.
It describes the correlation between two points in space-time,
\begin{eqnarray}    \label{g1_rt1}
g^{(1)}({\bf r},t;{\bf r}',t') = \langle \frac{\psi^{\dag}({\bf r},t) \psi({\bf r}',t')}{\sqrt{n({\bf r},t) n({\bf r}',t')}} \rangle ,
\end{eqnarray}
where $n ({\bf r},t)$ is the density of the quantum fluid.
In the density-phase representation, we have
\begin{eqnarray}    \label{madron1}
\psi({\bf r},t) = \sqrt{n ({\bf r},t)} e^{i \theta ({\bf r},t)} = \sqrt{n ({\bf r},t)} e^{i [\bar{\theta} + \delta \theta ({\bf r},t) ]} .
\end{eqnarray}
Thus the coherence function can be expressed as
\begin{eqnarray}    \label{g1_rt4}
g^{(1)}({\bf r},t;{\bf r}',t')
 &=& \langle e^{-i [ \delta \theta ({\bf r},t) - \delta \theta ({\bf r}',t') ]} \rangle
  \nonumber \\
 &=& e^{-{1 \over 2} \langle [\delta \theta ({\bf r},t) - \delta \theta ({\bf r}',t') ]^2 \rangle} ,
\end{eqnarray}
where $\bar{\theta}$ is the average value of the phase, and $\delta \theta ({\bf r},t)$ is the phase fluctuation at position $\bf r$ and time $t$.
The second expression in Eq.~\eqref{g1_rt4} is obtained by a standard second-order cumulant expansion.
As already known, in the 2D systems, there exists no real ordering except for a possible quasi-long-range one.
Under the Bogoliubov linear approximation, it can be derived that the equal-time coherence function is characterized by a power-law decay,
\begin{eqnarray}    \label{g1_rt5}
g^{(1)}({\bf r},{\bf r}') \approx |{\bf r} - {\bf r}'|^{- \eta} ,
\end{eqnarray}
where
\begin{eqnarray}    \label{eta1}
\eta = \frac{m D (\mu^2 + \Gamma^2)}{\pi \mu \Gamma} ,
\end{eqnarray}
is the decay exponent of the quasi-long-range correlation.

As already noted in the earlier work, the fluctuation of density is very weak \cite{Gladilin-2014}, hence the density distribution $n({\bf r},t)$ can be regarded as a constant.
In this way, the coherence function Eq.~\eqref{g1_rt1} can be rewritten as
\begin{eqnarray}    \label{g1_rt6}
g^{(1)}({\bf r},t;{\bf r}',t') \cong {1 \over n_0} \langle \psi^{\dag}({\bf r},t) \psi({\bf r}',t') \rangle ,
\end{eqnarray}
where $n_0$ is the average value of the density.
Related with $g^{(1)}$, another quantity we are concerned is the correlation of the $\delta \psi$, i.e. the fluctuation of $\psi$, defined as
\begin{eqnarray}    \label{g1_rt7}
G({\bf r},t;{\bf r}',t') \equiv \langle \delta \psi^{\dag}({\bf r},t) \delta \psi({\bf r}',t') \rangle .
\end{eqnarray}
The Fourier transform of these correlation functions are,
\begin{eqnarray}
    \label{g1_kt4}
g_{\bf k}^{(1)}(t,t') & \equiv & \langle \psi_{\bf k}^{\dag}(t) \psi_{\bf k} (t') \rangle
    \nonumber \\
    &=& {n_0 \over (2\pi)^2} \int d{\bf r} d{\bf r}' g^{(1)}({\bf r},t;{\bf r}',t') e^{-i k ({\bf r}-{\bf r}')} ,
    \\
    \label{g1_kt5}
G_{\bf k}(t,t') & \equiv & \langle \delta \psi_{\bf k}^{\dag}(t) \delta \psi_{\bf k} (t') \rangle
    \nonumber \\
    &=& {n_0 \over (2\pi)^2} \int d{\bf r} d{\bf r}' G({\bf r},t;{\bf r}',t') e^{-i k ({\bf r}-{\bf r}')} .
\end{eqnarray}
It is straightforward to show that they are related by
$g_{\bf k}^{(1)} (t,t') = (2 \pi)^2 n_0 \delta_{\bf k,0} + G_{\bf k} (t,t')$.
In our earlier paper \cite{Ji-2015}, by using the Bogoliubov linear approximation, it has been derived
\begin{eqnarray}    \label{g1_kt3}
& & G_{\bf k} (t,t') = (2 \pi)^2 D e^{-\Gamma (t-t')}
  \nonumber \\
& & \times \left\{
\begin{aligned}
& \left[ {1 \over \omega_{\bf k}} \left( {\mu^2 + \Gamma^2 \over E_{\bf k}^2} +i {\epsilon_{\bf k} + \mu \over \Gamma} \right) \sinh \omega_{\bf k} (t-t') \right.
\\
& \left. + {1 \over \Gamma} \left( 1+ {\mu^2 + \Gamma^2 \over E_{\bf k}^2} \right) \cosh \omega_{\bf k} (t-t') \right]
\text{,~for~} |{\bf k}| \leq k_c ,
\\
& \left[ {1 \over \omega_{\bf k}} \left( {\mu^2 + \Gamma^2 \over E_{\bf k}^2} +i {\epsilon_{\bf k} + \mu \over \Gamma} \right) \sin \omega_{\bf k} (t-t') \right.
\\
& \left. + {1 \over \Gamma} \left( 1+ {\mu^2 + \Gamma^2 \over E_{\bf k}^2} \right) \cos \omega_{\bf k} (t-t') \right]
\text{,~for~} |{\bf k}| > k_c ,
\end{aligned}
\right.
\end{eqnarray}
where $k_c=\sqrt{2m} \left( \sqrt{\mu^2+\Gamma^2} - \mu \right)^{1/4}$ is the critical momentum for bifurcation.
In particular, when $t=t'$, the above expressions reduce to the density fluctuation in the Fourier space,
\begin{eqnarray}    \label{nk}
\delta n_{\bf k} \equiv \langle \delta \psi_{\bf k}^{\dag}(t) \delta \psi_{\bf k}(t) \rangle
    = {D \over \Gamma} \left[ 1+ \frac{4 m^2 (\mu^2 + \Gamma^2)}{k^2 (k^2 + 4m \mu)} \right] .
\end{eqnarray}
Analogous to the one-dimensional (1D) case, in 2D systems, Eqs.~\eqref{g1_kt3} and \eqref{nk} are valid only when the nonlinear effect is weak.
In the next section, we will see they break down in the KPZ regime.

By introducing a two-point correlation function of the phase fluctuation \cite{Canet-2010},
\begin{eqnarray}    \label{cor1}
C({\bf r},t;{\bf r}',t') \equiv \langle [ \delta \theta ({\bf r},t) - \delta \theta ({\bf r}',t') ]^2 \rangle ,
\end{eqnarray}
we can rewrite the coherence function Eq.~\eqref{g1_rt4} as
\begin{eqnarray}    \label{g1_rt3}
g^{(1)} ({\bf r},t;{\bf r}',t') = e^{- {1 \over 2} C ({\bf r},t;{\bf r}',t')} .
\end{eqnarray}
We note that $C({\bf r},t;{\bf r}',t')$ is an important function to characterize the KPZ dynamic scaling.
At long time and large spatial scale, the KPZ dynamic scaling is represented through an asymptotic form of the correlation function,
\begin{eqnarray}    \label{cor2}
C({\bf r},t;{\bf r}',t') = |{\bf r} - {\bf r}'|^{2 \chi} f(|t-t'| / |{\bf r} - {\bf r}'|^z) ,
\end{eqnarray}
where $f(|t-t'| / |{\bf r} - {\bf r}'|^z)$ is the KPZ scaling function, and $\chi$ and $z$ are two characteristic exponents.
For 2D systems, $\chi = 0.4$ and $z = 1.6$ \cite{Kim-1989}.
Substituting Eq.~\eqref{cor2} into Eq.~\eqref{g1_rt3}, we find
\begin{eqnarray}    \label{scale1}
f(|t-t'| / |{\bf r} - {\bf r}'|^z) = - \frac{2 \ln g^{(1)}({\bf r},t;{\bf r}',t')}{|{\bf r} - {\bf r}'|^{2 \chi}} .
\end{eqnarray}
Eq.~\eqref{scale1} suggests that by evaluating the coherence function $g^{(1)}$, the scaling function $f$ can be determined.
If we set $t=t'$, Eq.~\eqref{scale1} leads to the static or equal-time scaling function,
\begin{eqnarray}    \label{scale2}
f(0) = - \frac{2 \ln g^{(1)}({\bf r},{\bf r}')}{|{\bf r} - {\bf r}'|^{2 \chi}} .
\end{eqnarray}
Eq.~\eqref{scale2} implies that in the KPZ regime, the static scaling function $f(0)$ is a spatial invariant.
Therefore, provided a large spatial distance, the right-hand-side of Eq.~\eqref{scale2} is expected to be independent of $\bf r$ and $\bf r'$.
We shall discuss the KPZ scaling behavior of the EP system in terms of this asymptotic form.


To reveal the KPZ scaling of spatiotemporal coherence for the EP system, we have performed numerical studies on stochastic GGPE in 2+1-dimension.
The stochastic GGPE \eqref{ggpe} is numerically solved using the splitting-flip method: the wave function evolves alternatively as $\psi_{\bf k} (t) \rightarrow e^{-i T \Delta t} \psi_{\bf k} (t)$ in the Fourier space and $\psi ({\bf r},t) \rightarrow e^{-i V \Delta t} \psi ({\bf r},t)$ in the real space, where $T = k^2 / (2m)$ and $V = g |\psi|^2 +i [ P_0 / (1+|\psi|^2/n_s) - \gamma ]$.
The two evolutions are connected by a fast Fourier transform (FFT), and the noise term is added every time when the real space wave function $\psi ({\bf r},t)$ is updated.
The simulations are conducted on square lattices with periodic boundary conditions.
We start from initial configurations with a uniform density and random local phases.
After about $10^5 \sim 10^6$ iterations with time step $\Delta t$=0.001, the system is stabilized at a steady state.
Then the ensemble averaging of coherence function $g^{(1)}({\bf r},t)$ is carried out over about 1000 samples in the time evolution sequence of $\psi ({\bf r},t)$ with 1000 iterations between every two neighboring samples to reduce their correlations.

The simulations on 2+1 dimensional systems are costly in computing time. In this work, we perform the simulations on the architecture with CUDA-enabled graphics processing unit (GPU), which is well-suited to implement the data-parallel computation.
The CUDA environment, including the cuFFT library, allows developers to use C as a more efficient programming language than on the conventional architecture of CPU \cite{cuda-2011}.
However, to use CUDA, data values must be transferred between the CPU and GPU repetitively.
These transfers are particularly time consuming and should be minimized in an iterative sampling procedure.
Therefore we perform all of the calculations on the GPU, including FFT operations and time evolutions, to avoid unnecessary data transfer.
Thanks to the acceleration by GPU, the performance is improved by approximately a factor of 10 in comparison with the CPU-based computation.


\section{Numerical results \label{sec:result}}

\subsection{Density fluctuation $\delta n_{\bf k}$}

\begin{figure}[htbp]
\centering
\includegraphics[width=0.45\textwidth]{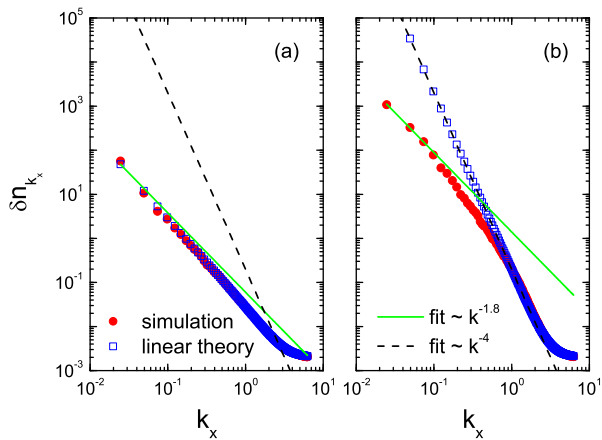}
\caption{(Color online) Density fluctuation  $\delta n_{\bf k}$ along the $k_x$ direction in 2D Fourier space of $512 \times 512$ systems for different interaction strengths: (a) $g=2$ and (b) $g=0$.
The red filled circles and blue hollow squares are results of noisy GGPE and Bogoliubov linear theory, respectively.
Two power-law dependences on $k_x$ are plotted by the black dashed and green solid lines as guide to the eyes.}
\label{fig:dnk}
\end{figure}

Figure~\ref{fig:dnk} presents our results of density fluctuation $\delta n_{\bf k}$ along the $k_x$-axis of 2D Fourier space of $512 \times 512$ systems for (a) $g=2$ and (b) $g=0$.
Here we set polariton mass $m = 1$, pump strength $P_0 = 20$, and damping rate $\gamma = P_0 / 2$.
The diffusion coefficient is fixed at an intermediate strength of $D = 0.01$, which is lower than the threshold to destroy the condensate or to create vortex spontaneously \cite{Gladilin-2019b}.
When the coupling $g$ is strong, the results of simulation (red filled circles) and Bogoliubov theory (blue hollow squares) agree very well in panel (a).
When $g = 0$, the simulations on GGPE show the characteristic $k_x^{-1.8}$-dependence of KPZ scaling in panel (b).
In constrast, a $k_x^{-4}$-dependence is obtained by the Bogoliubov linear theory.
This discrepancy is attributed to the omission of nonlinear terms in the Bogoliubov linear theory, and is similar to the situation in 1D \cite{Gladilin-2014}.
Here one notices that the linear theory breaks down for $k_x \le 1$ or so, as seen in Fig.~\ref{fig:dnk}(b), indicating that the KPZ nonlinear effect dominates the small momentum regime.

\subsection{Equal-time coherence function $g_1 ({\bf r})$}

\begin{figure}[htbp]
\centering
\includegraphics[width=0.45\textwidth]{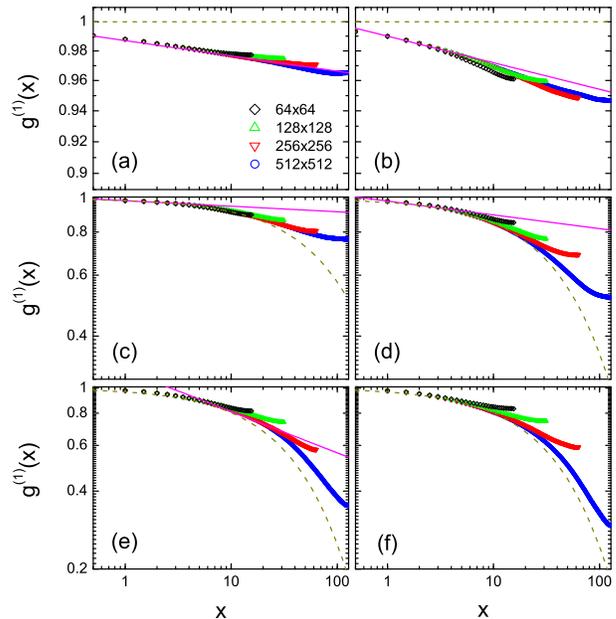}
\caption{(Color online) Scaling analysis of  $g^{(1)} (x)$ for different interaction strength: (a) $g=2$, (b) $g=1$, (c) $g=0.5$, (d) $g=0.2$, (e) $g=0.1$, and (f) $g=0$, presented in double logarithmic scales.
The discrete symbols of different color correspond to the numerical results from different system size.
The magenta lines denote asymptotic $g_{\infty}^{(1)} (x)$ with quasi long range orders.
There is no magenta line in panels (f) as Eq.~\eqref{eta1} diverges at $g=0$.
The grey dashed curves are the asymptotic $g_{\infty}^{(1)} (x)$ of KPZ stretched exponential decay.}
\label{fig:g1}
\end{figure}

Figure~\ref{fig:g1} illustrates the equal-time first order coherence function along $x$-axis, i.e. $g^{(1)} (x)$ [$\equiv g^{(1)} ({\bf r,r}') |_{x'=y=y'=0}$], on the $64 \times 64$, $128 \times 128$, $256 \times 256$, and $512 \times 512$ systems for different interaction strength $g$ in double logarithmic scales.
In each panel, the grey dashed curve shows the asymptotic coherence function $g_{\infty}^{(1)} (x)$ of infinite system.
Here, $g_{\infty}^{(1)} (x)$ is derived from [cf. Eq.~\eqref{scale2}]
\begin{eqnarray}    \label{scale3}
g_{\infty}^{(1)} (x) = \exp \left[ -f_{\infty}(0) x^{2 \chi} /2 \right]  ,
\end{eqnarray}
where $f_{\infty}(0)$ means the static scaling function in the thermodynamic limit.
Because of its spatial invariant nature, $f_{\infty}(0)$ is extrapolated to be an $x$-independent constant.
For $g>0.5$, $f_{\infty}(0) \approx 0$ in panels (a) and (b).
While for $g \le 0.5$, $f_{\infty}(0)$ are nonzero constants in panels (c)-(d).
We shall discuss the scaling analysis and extrapolation of $f_{\infty}(0)$ in more details in Sec.~III.E.
In Fig.~\ref{fig:g1}, we also show results of Bogoliubov theory (magenta solid lines) for comparison.

In Fig.~\ref{fig:g1}, the symbols represent the numerical results of $g^{(1)} (x)$ for different system sizes, the magenta solid curves show fittings to the power law decay $g_{\infty}^{(1)} (x) \sim |x|^{- \eta}$, and the grey dashed curves are $g_{\infty}^{(1)} (x)$ determined from Eq.~\eqref{scale3}. 
With the decrease of interaction $g$, the coherence function $g^{(1)} (x)$ deviates from the power-law decay (magenta curve) and approaches to the KPZ stretched exponential decay (grey dashed curves) $\sim \exp⁡ [- (x / l)^{0.8} ]$, where $l$ can be interpreted as a characteristic coherence length.
This crossover is consistent with the tendency in Fig.~\ref{fig:dnk}, that is the correlation in the small momentum regime (or large spatial scale) is dominated by the KPZ nonlinearity, while the properties in the large momentum regime (or short spatial range) can still described by the Bogoliubov theory.
Moreover, by comparing the numerical results of small and big systems in Fig.~\ref{fig:g1}, one notices that the finite size effect is quite severe in the KPZ regime but not in the linear regime.
For example, in panels (d)-(f), the correlation at longer distance shown by blue circles cannot be extrapolated from those of short distance plotted by black diamonds.
This size dependence is not seen in panels (a) and (b), implying that numerical simulations without a proper scaling analysis can hardly figure out the real KPZ scaling behaviors.

\subsection{Vortex effect on the spatial coherence}

Before we study the KPZ scaling properties, a discussion about the vortex excitation in the EP system is in order.
Vortex is a topological defect manifested as a suppression of fluid density in 2D plane with circular flow of phase gradient around its empty center.
When the free vortices and anti-vortices get bounded and form vortex pairs, a topological phase transition takes place.
The appearance of the vortices as well as the topological order significantly modifies the phase distribution of the quantum fluid.
Therefore it is important to clarify the vortex effect on the spatial correlation in connection with the KPZ nonlinearity.

\begin{figure}[htbp]
\centering
\includegraphics[width=0.45\textwidth]{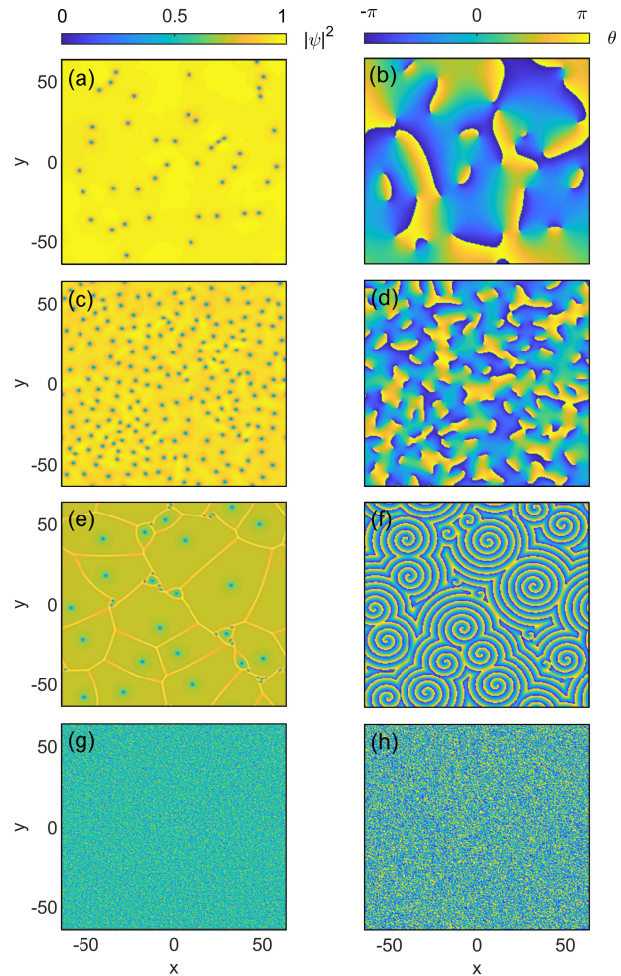}
\caption{(Color online) Representative snapshots of vortex patterns for different $P_0$:
(a) and (b) attractive regime for $P_0=0.02$, (c) and (d) metastable regime for $P_0=2$, (e) and (f) complex excitation regime for $P_0=10$, (g) and (h) spatiotemporal chaotic regime for $P_0=22$, simulated on $256 \times 256$ lattices with coupling constant $g=1$.
The left and right panels represent the density and phase distributions, respectively.
The snapshots are taken after long time evolutions from the initial states of 128 vortex pairs embedded randomly in homogeneous quantum fluids.}
\label{fig:Multivortex}
\end{figure}

The vortex dynamics in the noiseless excitons-polariton fluids has already been investigated in Refs.~\cite{Gladilin-2017, Gladilin-2019a}.
It is found that a vortex has not only the tangential flows but also the radial ones, which increase with the pumping strength $P_0$.
As a consequence, the current distributions around a vortex evolve from concentric rings to spiral flows with the increase of $P_0$.
Such behaviors are very different from the quantum fluid at equilibrium, leading to distinctive nonequilibrium vortex patterns.
As represented in Fig.~\ref{fig:Multivortex} by the snapshots of wave functions, the vortex patterns can be roughly classified into four regimes depending on the control parameter $P_0$ (here the interaction strength is fixed at a typical value of $g=1$):

(1) When $P_0$ is very small, the radial flow is negligible.
In this case, the attraction between the vortex and anti-vortex is quite strong.
So the unlike vortices easily annihilate each other when they get close.
For this reason, the vortex pair in a nonequilibrium quantum fluid has only a short lifetime, and the vortex density of the system is always at a low level [see in Figs.~\ref{fig:Multivortex}(a) and \ref{fig:Multivortex}(b)].

(2) With the increase of $P_0$, the radial flows of both vortex and anti-vortex grow gradually, which counteract the attraction between them and also suppress the tendency of annihilation.
If $P_0$ is large enough, the attraction between the vortex pairs can be replaced by a repulsion.
Thus, the vortex and anti-vortex pairs are decoupled, and they move separately as if they are of the same chirality.
As a result, the system reaches a metastable state where high density vortices can exist for a rather long time [see in Figs.~\ref{fig:Multivortex}(c) and \ref{fig:Multivortex}(d)].

(3) Increasing $P_0$ further, some complex excitations like the domain walls and even new vortices can be generated.
In this case, vortex cells or clusters appear in the density distribution [see in Fig.~\ref{fig:Multivortex}(e)], and meanwhile spiral waves appear in the phase space [see in Fig.~\ref{fig:Multivortex}(f)].
Although here $P_0$ might be strong enough to incur vortex nucleation, the vortex number fluctuates only locally and does not lead to a global instability.

(4) For even larger $P_0$, the system enters a spatiotemporal chaotic regime, where an explosive increment of the vortex number can be triggered by adding vortices, and the whole fluid becomes turbulent [see in Figs.~\ref{fig:Multivortex}(g) and \ref{fig:Multivortex}(h)].

\begin{figure}[htbp]
\centering
\includegraphics[width=0.45\textwidth]{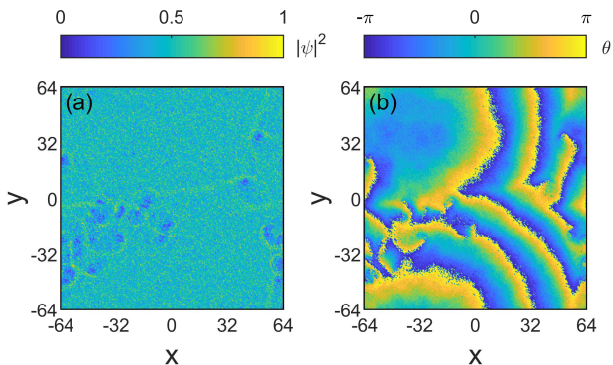}
\caption{(Color online) Snapshot of a steady state configuration with many vortices simulated on a $256 \times 256$ lattice: (a) density, (b) phase.
The pumping strength $P_0=2$, coupling constant $g=0$.}
\label{fig:Vortex}
\end{figure}

Now we look into the vortices of noisy systems.
As have been demonstrated in the previous section, the KPZ effect tends to be dominant in the weak interaction regime of the EP systems.
For this reason, we confine our discussion in the $g=0$ case in the rest part of this section.
When $g$ is vanishing, we find that among the above-mentioned four regimes of vortex patterns, only (3) and (4) are realizable, while (1) and (2) are irrelevant.
The reason is that in order to produce a condensate in the free particle limit of $g=0$, the spatiotemporal correlation can only be developed through applying an external pumping with a sufficient strength.
Moreover, since we now consider a noisy system, it also requires a strong pumping to stabilize the condensate against the noise.
Based on our simulations with the noisy GGPE, we find that when $g=0$, condensation appears for $P_0 \ge 2$.
In fact, this is already strong enough to drive the system into a regime that features complex excitations.
In Fig.~\ref{fig:Vortex}, we show a snapshot of the simulated wave function at $P_0=2$ on a $256 \times 256$ lattice.
The initial state for the simulation consists of a single vortex-anti-vortex pair embedded in a homogeneous fluid.
After a process of relaxation with vortex proliferations, as can be seen in Fig.~\ref{fig:Vortex}, the system approaches a steady state characterized by clusters of spontaneously generated vortices and domain walls.

\begin{figure}[htbp]
\centering
\includegraphics[width=0.36\textwidth]{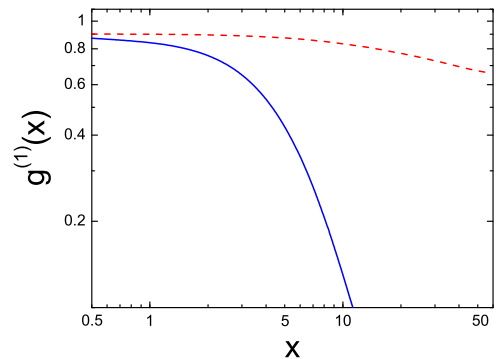}
\caption{(Color online) Time-averaged first order static coherence function along the $x$-axis.
The blue solid curve corresponds to a system with many spontaneous vortices,
and the red dashed curve corresponds to a system with no vortex added.
The parameters are the same as in Fig.~\ref{fig:Vortex}.}
\label{fig:g1_Vortex}
\end{figure}

In Fig.~\ref{fig:g1_Vortex}, we calculate the time-averaged $g^{(1)} (x)$ in the double logarithmic scale for this system.
The blue solid curve corresponds to the situation of many spontaneous vortices represented by Fig.~\ref{fig:Vortex}, and the red dashed curve corresponds to a situation of the same parameters but with no vortex added.
Apparently, the red dashed curve features two length scales with a crossover around $x=7$.
As has been addressed in Sec.~III.B, the coherence within a short distance is determined by a quasi-long-range order, while the long distance behavior is subjected to the KPZ scaling.
In comparison, the blue solid curve show a fast decay with the distance, indicating that the spatial coherence is heavily suppressed with the introduction of vortices, though a quasi-long-range order may still exist in a very short scale up to a characteristic distance between the vortices \cite{Wachtel-2016}.

We have also tested the vortex evolutions at larger $P_0$.
It is found that when $P_0 \ge 8$, a large amount of vortices can be generated shortly after a vortex pair is added into a homogeneous fluid.
The evolution ends up with a state in the spatiotemporal chaotic regime composed of lots of small vortices occupying the whole space uniformly.
For such a turbulent fluid, the spatial coherence declines even more dramatically (not shown here).

\subsection{Dynamic scaling function $f(t/r^z)$}

In this section, we examine the KPZ scaling properties of EP system in terms of dynamic scaling function $f(t/r^z)$ defined in Eq.~\eqref{scale1}.
To enhance the KPZ nonlinear effect, we shall apply a strong pumping of $P_0=20$ here.
As noted in the last section, this pumping strength corresponds to a spatiotemporal chaotic regime in the presence of vortices.
Its turbulent nature destructs the spatiotemporal coherence completely.
Allowing for this, here we shall only be concerned with the irrotational fluids without any vortex.

\begin{figure}[htbp]
\centering
\includegraphics[width=0.45\textwidth]{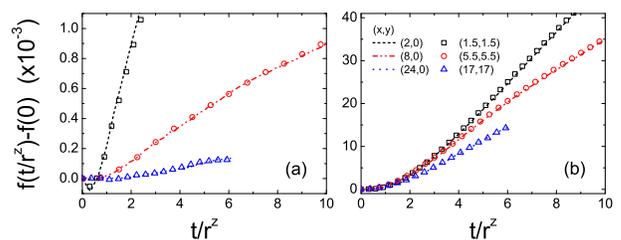}
\caption{(Color online) Dynamic scaling functions $f(t/r^z)$ calculated at three sets of points for $256 \times 256$ systems.
In each set, the curve and symbol are calculated at roughly the same distance $r$ from $(0,0)$ in 2D plane.
The curves are obtained along the $x$-axis, and the symbols along the $y=x$ line.
Panel (a) is for $g=2$, and (b) for $g=0$.}
\label{fig:f}
\end{figure} 

Figure~\ref{fig:f} plots our numerical results of the shifted dynamic scaling function $f(t/r^z)-f(0)$ obtained from the numerical simulations on noisy GGPE.
In each panel, we present three sets of data, with each set including two points having roughly the same distance $r$ from $(0,0)$ in 2D plane.
One of the points is on the $x$-axis (represented by curves), and the other along the diagonal line $y=x$ (denoted by symbols).
It can be seen that the dynamic scaling functions are identical for the points of the same distance $r$.
Meanwhile, the functional form changes with the interaction strength $g$ as well as distance $r$.

\begin{figure}[htbp]
\centering
\includegraphics[width=0.45\textwidth]{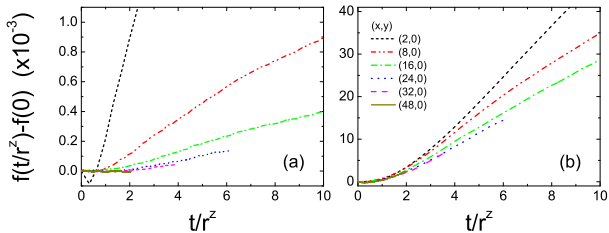}
\caption{(Color online) Dynamic scaling functions $f(t/r^z)$ calculated at six different points in the direction of $x$-axis for $256 \times 256$ systems.
Panel (a) is for $g=2$, and (b) for $g=0$.}
\label{fig:Scaling}
\end{figure} 

Figure~\ref{fig:Scaling} shows the shifted dynamic scaling function $f(t/r^z)-f(0)$ at six different points along the $x$-axis for $256 \times 256$ systems.
Here one finds that for the short distance $r<24$, the function $f$ has different forms depending on the distance $r$.
While for the large distance $r \ge 24$, all the function $f$ collapse to the same form.
This collapse becomes even more clear when the interaction $g$ approaches to zero in panel (b), signifying the dominance of KPZ type correlation at large spatial scale.

\subsection{Static scaling function $f(0)$}

\begin{figure}[htbp]
\centering
\includegraphics[width=0.45\textwidth]{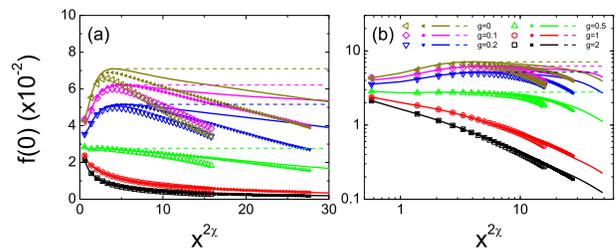}
\caption{(Color online) Static scaling function $f(0)$ calculated along the $x$-axis for different interaction strength $g$: (a) in linear scale, (b) in double logrithmic scale.
The big symbols, small symbols, and solid curves are obtained from $128 \times 128$, $256 \times 256$, $512 \times 512$ systems respectively.
The horizontal dashed lines are extrapolations to the systems of infinite size.}
\label{fig:f0}
\end{figure}

In order to better understand the different scaling properties between the strong and weak coupling cases associated with the KPZ universality class, we carry out a study on the static scaling function $f(0)$ introduced above in Eq.~\eqref{scale2}.
In Fig.~\ref{fig:f0}, we show $f(0)$ along the $x$-axis for different $g$, where the big symbols, small symbols, and solid lines are calculated from $128 \times 128$, $256 \times 256$, $512 \times 512$ systems, respectively.
Panels (a) and (b) present the same results in linear and double logarithmic scales, respectively.
In this graph, one can readily identify two different trends of $x$-dependence with the change of system size:

(1) When $g<0.5$, $f(0)$ first increases and then decreases with the increase of $x$, with a maximum located at around $x=5$.
In the uphill region of $x<5$, $f(0)$ only changes a little for different system sizes.
On the contrary, for the downhill part of $x>5$, $f(0)$ varies with the system size and becomes flatter for increasing system size and should tend to a constant for $x \rightarrow \infty$.
This asymptotic behavior corroborates our previous statement about the space-independent nature of Eq.~\eqref{scale2}, and hence allows us to extrapolate $f_{\infty} (0)$ as the horizontal dashed lines in Fig.~\ref{fig:f0}, which has already been used in deriving $g_{\infty}^{(1)} (x)$ in Fig.~\ref{fig:g1}.
Apparently, $f_{\infty} (0)$ converges to a non-zero value which increases with decreasing $g$.
This confirms that with the decrease of $g$, the KPZ scaling effect plays a more important role in the spatiotemporal correlation.

(2) When $g \ge 0.5$, $f(0)$ monotonically declines with the increase of $x$.
For a given $g$, the results of $f(0)$ obtained from different system sizes all have the same form, i.e. they almost coincide with each other, showing negligible size dependence.
From panel (b) it can be clearly seen that for $g \ge 0.5$, all $f(0)$ approach to zero value when $x \rightarrow \infty$ as can be expected from Eq.~\eqref{g1_rt5}.
This means the KPZ scaling effect tends to vanish for the large enough systems.
Accordingly, we can extrapolate $f_{\infty} (0)=0$ in the present case.

Based on these discussions, as well as the results presented in Sec.~III.B, one can see that the finite size effect deserves careful analyses in the KPZ regime.
It is difficult to extract large scale properties directly from the results of small systems.
A proper multiple scaling analysis is crucial for understanding the scaling features of KPZ universality class.

\section{Conclusions \label{sec:con}}

We perform a numerical study on the scaling properties of the spatial and temporal coherence of 2D EP systems in the incoherent pumping regime.
Based on the simulations with noisy GGPE, we show that both the static and dynamic coherence functions display the scaling features of KPZ universality class, especially in the weak interaction regime.
By comparing the numerical results of finite systems with the asymptotic scaling functions of infinite size, we find that the quasi long range order persists only in a short distance, and the KPZ dynamical scaling behavior is overwhelming at large scales.
We also find that when the interaction is weak, the vortex state is stabilized in a regime with complex excitations.
In the KPZ regime, the vortex pairing effect is suppressed due to repulsive interaction, while introducing vortex excitations can drive the system into a state close to spatiotemporal chaos.

\begin{acknowledgments}
The authors thank V. N. Gladilin, J.-F. Yu, P. Ao, and J. Bloch for useful discussions.
K.J. was supported by the Shanghai Pujiang Program (Project No. 17PJ1407400).
\end{acknowledgments}

\end{document}